\documentclass[prb,floatfix,reprint,twocolumn,superscriptaddress,showpacs,titlepage,amssymb,amsmath]{revtex4-1}

\usepackage{graphicx, times, epsfig, color, float, mathtools}
\usepackage{subfigure}
\usepackage{xcolor}
\usepackage[colorlinks = true, linkcolor = blue, urlcolor  = blue, citecolor = blue, anchorcolor = blue]{hyperref} 

\def\be{\begin{equation}}
\def\ee{\end{equation}}
\def\bea{\begin{eqnarray}}
\def\eea{\end{eqnarray}}
\def\etal{{\it et al.}}
\def\deg{$^{\circ}$}

\def\PDNI{Ni$_{40}$Pd$_{40}$P$_{20}$}

\begin{document} 
\title{First principles modeling of the structural, electronic, 
and vibrational properties of Ni$_{40}$Pd$_{40}$P$_{20}$ bulk metallic glass}
\author{Raymond Atta-Fynn} 
\email{attafynn@uta.edu}
\affiliation{Department of Physics, University of Texas at Arlington, Arlington, TX 76019}

\author{David A. Drabold}
\email{drabold@ohio.edu}
\affiliation{Department of Physics and Astronomy, Ohio University, Athens, Ohio 45701}

\author{Parthapratim Biswas}
\email{partha.biswas@usm.edu}
\affiliation{ Department of Physics and Astronomy, The University of Southern 
Mississippi, Hattiesburg, MS 39406}

\begin{abstract}
The structural, vibrational, and electronic properties of 
{\PDNI} bulk metallic glass have been 
studied using {\it ab initio} molecular-dynamics simulations 
and total-energy optimization. Structural analyses of the 
resulting {\it ab initio} models show the presence 
of a few to no P-P bonds and two main building blocks, 
consisting of tricapped trigonal prism (TTP) and capped 
square anti-prism (CSAP) with P as the center of these 
blocks.  The computed Pd and Ni {\it K}-edge spectra of extended 
x-ray absorption fine structure (EXAFS) are found to be 
in good agreement with experimental data.  The configurational 
averaged static structure factor and the generalized 
vibrational density of states are also observed to be 
in good agreement with experimental data.  
\end{abstract}

\maketitle 
\section{Introduction}

Metallic glasses (MG) are a class of amorphous metallic alloys that are produced by
rapid quenching from a molten state at a sufficiently fast cooling
rate so that crystal nucleation and growth can be avoided. The ability
of a metal alloy to form a glass -- the so called glass forming
ability (GFA) -- depends on the critical cooling rate, which is
defined as the {\it slowest cooling rate} at which the melt can be
quenched into a glassy state. The smaller the critical cooling rate,
the higher the GFA of a metallic alloy.\cite{Jafary2018} Another
measure for the GFA of an alloy is the Turnbull criterion.\cite{Turnbull1949,Turnbull1969}
The latter predicts that glass formation should occur if the
{\it reduced} glass transition temperature, $T_{{\rm r}g}$,
satisfies the condition $T_{{\rm r}g}>2/3$, where $T_{{\rm r}g}$
is defined as $T_{{\rm r}g}=T_g/T_m$, with $T_g$ and $T_m$
being the glass transition and melting temperatures of the alloy,
respectively. The formation of metallic glasses is a difficult
process in comparison with non-metallic glasses, such as silicates, ceramic, and
polymeric glasses, because pure metals and ordinary metallic
alloys readily crystallize when quenched from a molten state.  The
first metallic glass, Au$_{75}$Si$_{25}$, was reported to be
obtained by rapid quenching from the liquid state
at a very high cooling rate of 10$^5$--10$^6$ K/s, with a critical thickness of
$\sim$ 50 $\mu{\rm m}$.\cite{Klement1960} In general, metallic
glasses formed via very high cooling
rates (10$^5$--10$^{10}$ K/s) have thicknesses in the
{\it sub-millimeter} range; such metallic glasses are known
as {\it conventional} metallic glasses (MG). By contrast, a
{\it bulk} metallic glass (BMG) is defined to be one
with a critical thickness of at least a millimeter -- the word
``bulk" being indicative of the millimeter length scale.\cite{Wang2004}
The first BMG was produced from a Pd-Cu-Si alloy by quenching from
the  melt at a relatively lower cooling rate of 10$^3$ K/s.

In comparison to their crystalline metallic counterparts,
BMGs possess unique properties such as higher strength, lower
Young's modulus, improved wear resistance, good fatigue
endurance, and excellent corrosion resistance,  due mainly
to the combination of metallic bonding and amorphous
structure.\cite{Ashby2006} As a result, they are extensively
employed in biomaterials research and they have potential
uses for improved corrosion resistance, biocompatibility,
strength, and longevity of biomedical implants, which make
BMGs as promising candidates for cardiovascular
applications.\cite{Jafary2018,Wang2004,Schroers2009,Li2016,Huang2015}
Bulk metallic glasses also find applications in the manufacture of
electromagnetic instruments, cell-phone cases, optical mirrors,
striking face plate in golf clubs, connecting parts
for optical fibers, soft magnetic high-frequency power coils, etc.\cite{Inoue2008} Currently, a wide
range of multi-component BMG alloys \cite{Wang2004,Inoue2000}
exists; however, the industrial production of BMGs remains a difficult
problem due to the requirement of high cooling rates
to produce a metallic glass with a thickness on
the macroscopic length scale.

{\PDNI} was first prepared by
Turnbull and co-workers,~\cite{Turnbull1982,Turnbull1984} who
employed a critical cooling rate of about 1.4 K/s and
the application of B$_2$O$_3$ fluxing method to purify the melt
and avoid heterogeneous nucleation.
The study indicates that a value of $T_{{\rm r}g}$ close to 2/3
was achieved in their experiments. In a later work, He {\it et al.\,}\cite{He1996}
produced {\PDNI} BMG using a critical cooling rate
of 10$^{-3}$ K/s.

The computational modeling of BMGs presents a significant
challenge due to high atomic coordination and the presence
of strong chemical order in the system. This is further
compounded by the heavy constituent atoms of BMGs, often
involving $d$ and $f$ electrons, which exacerbate the
problem by increasing the computational complexity of
density-functional calculations.  Furthermore, the
quantum-mechanical computations of the energies and forces
are hampered by the existence of finite electronic density
of states at the Fermi level, which requires extra care
for computation of the density matrix and a sophisticated
mixing scheme for self-consistent field calculations of
energies and forces.

In this paper, we address  the structural, electronic, and vibrational properties
of {\PDNI} bulk metallic glass using {\it ab initio} molecular-dynamics simulations
based on density-functional theory by employing a bigger system
size and relatively longer cooling rates. Thus far, only a few {\it ab initio}
studies on the electronic and structural properties of {\PDNI}
have been reported in literature. \cite{Takeuchi2007,Kumar2011,Guan2012,Reyes2015}
The scope of the present study goes much further by addressing the vibrational properties
of glassy {\PDNI}, starting with a notably larger model and a relatively
slower cooling rate than the ones that were used in earlier studies.
This was addressed by examining the phonon/vibrational density of states of the
constituent atoms and it was further validated by making comparisons with available
experimental data.  Since the vibrational properties of a solid are more sensitive to the
local structure and composition than their electronic counterpart, it is of crucial
importance in the glass-structure determination. The degree of distortion of the
local structural motif(s) can readily manifest in the vibrational excitation spectra,
given the miniscule amount of energy (with a few tens of meV) needed to excite
vibrational motion.  More importantly, unlike earlier studies where the results are solely based on
a single atomic configuration, the present study is the first of its kind
where the effects of configuration averaging on the physical observables have
been discussed at length.  Given the importance of configuration fluctuations
in small multi-component glassy/disordered systems, which are characterized by
several distinctly different bonding environments, an averaging over
independent configurations is essential to yield statistically reproducible results on
structural, electronic, and vibrational properties of glassy systems. To this end,
the results presented here were averaged over ten (10) independent
structural configurations to minimize the effects of configuration fluctuations.

The layout of the paper is as follows. In sec. II, we present the 
computational methodology that involves density functional theory 
based {\it ab initio} molecular-dynamics simulations, followed by
total-energy relaxation. The results and associated discussions from our
simulations are presented in sec. III,
with an emphasis on the electronic, structural, and vibrational
properties of the glass. Finally the summary and conclusions of
our work are presented in sec. IV.

\section{Computational Methodology}
{\it Ab initio} molecular-dynamics (AIMD) simulations and 
subsequent geometry relaxations were performed using the 
density functional theory code {\sc Siesta}.~\cite{siesta} 
{\sc Siesta} employs local basis sets, and the electron-ion interactions 
were described using norm-conserving 
Troullier-Martins pseudopotentials,\cite{tm} factorized in the 
Kleinman-Bylander form.\cite{kb} The Perdew-Zunger 
parameterization\cite{Ceperley1980,Perdew1981} of the local density 
approximation (LDA) to the exchange-correlation energy was employed. 
Single-$\zeta$ (SZ) basis functions were used for the finite 
temperature molecular-dynamics simulation, while double-$\zeta$ 
with polarization (DZP) basis functions were used for 
geometry relaxations (i.e., total-energy minimizations). 
Only the $\Gamma$-point, $\vec{k}$={\bf 0}, was used to perform 
the Brillouin zone integration.

The {\PDNI} simulations commenced with the generation of 
ten (10) independent initial random configurations, each 
comprising $N=300$ atoms (120 Ni atoms, 120 Pd atoms, and 
60 P atoms) placed in a cubic supercell at the experimental 
density~\cite{Haruyama2007} of 9.4 g/cm$^3$. Each configuration 
was subjected to independent NVT runs by integrating the 
equations of motion using the velocity-Verlet algorithm, with 
a time step of $\Delta t=1$ fs. The temperature of the system was 
controlled by a chain of Nos\'{e}-Hoover thermostats.~\cite{chains,hoover,nose}  
The AIMD simulations commenced by equilibrating each system at 2400 K for 20 ps. 
After equilibration, each system was quenched to 300 K over a period of 200 ps; 
this corresponds to an average cooling rate of 10.5 K/ps 
(or 1.05$\times 10^{11}$ K/s).  
The geometry at the end of the 300 K dynamics was optimized by 
minimizing the total energy using a conjugate-gradient algorithm. 
The final relaxed geometry was used for structural analysis.  
The convergence criterion for the geometry 
optimization was set to a maximum force of 0.01 eV/{\AA} on each atom.

The vibrational density of states of each of the ten (10) 
geometrically-relaxed configurations was computed within the 
harmonic approximation.  Small atomic displacements of 
0.005 {\AA} along six coordinate directions 
($\pm x$, $\pm y$, $\pm z$) were applied to compute the total 
force on each atom. This corresponds to a total of 6$N$ force 
computations, where $N$ is the number of atoms.  The dynamical 
matrix was constructed from the numerical derivatives of the 
forces with respect to the atomic displacement and diagonalized 
to obtain the vibrational eigenfrequencies and eigenvectors. 
The procedure was repeated for each of ten (10) configurations. 
All the properties presented in the sections to follow were 
averaged over the ten (10) independent configurations.

\section{Results and Discussion}

\subsection{Structural Properties}
\begin{figure}[t!]
\includegraphics[width=0.75\linewidth]{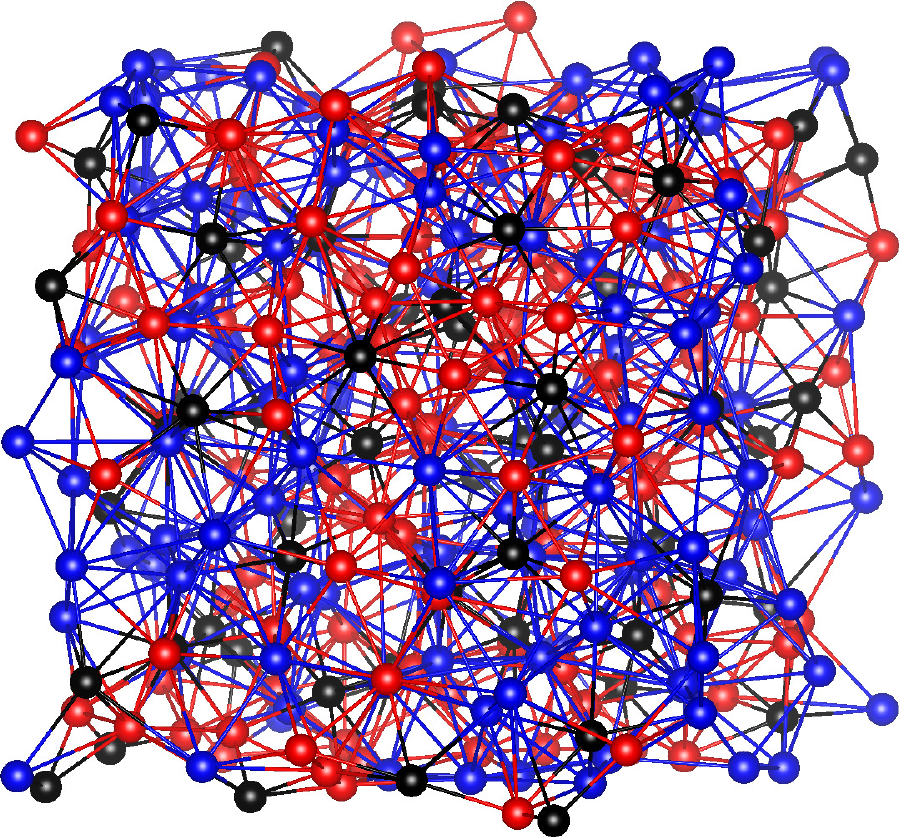}
\caption{\label{ball_stick}
(Color online)
A ball-and-stick representation of a 300-atom model of 
{\PDNI}. Pd, Ni, and P atoms are shown in blue, red, 
and black colors, respectively.
}
\end{figure}

\begin{figure}[t!]
\includegraphics[clip,width=0.75\linewidth]{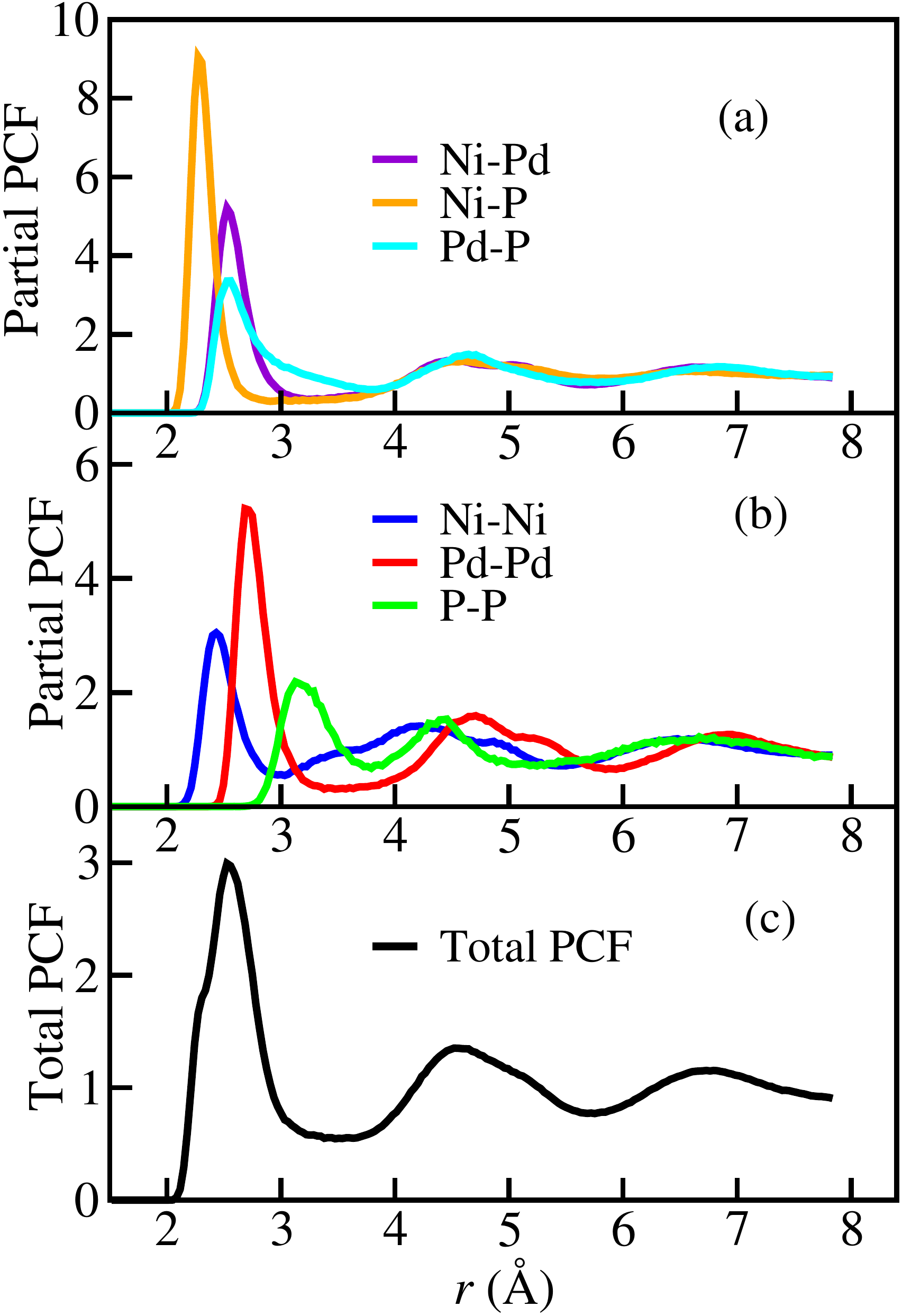}
\caption{\label{rdfplot}
(Color online)
Total and partial pair-correlation functions (PCF) computed from a 300-atom 
model of {\PDNI} glass. Panels (a) and (b) show various pair-correlation 
functions, whereas panel (c) depicts the total pair-correlation function.  
The results presented here were obtained by averaging over ten (10) independent 
configurations. 
}
\end{figure}

Figure \ref{ball_stick} depicts a ball-and-stick model of the optimized 
atomic structure of {\PDNI}. The continuous-random nature of the network 
is evident from the distribution of atoms in the network. 
Due to the dense random nature of the network, 
atoms can form bonds with a high number 
of neighboring atoms, resulting in a high coordination number 
in the range of 7-17 (cf. the values of $N$ in Table~\ref{TABLE2}). 
The configurationally averaged total and partial pair-correlation 
functions (PCF) are shown in Fig.\,\ref{rdfplot}. 

While the total PCF exhibits a first minimum near 3.5 {\AA}, the 
partial PCFs show varying first shell minima, from 2.3 {\AA} 
for Ni--P to 3.8 {\AA} for Pd--P and P--P. A direct implication 
of this variation is that there are virtually no or very little P--P bonds in 
the network.  
This is evident from the pair-correlation data in 
Fig.\,\ref{rdfplot}(b); the P--P correlation begins 
from just under 3.0 {\AA} but a significant number 
of P atoms form a bond with Ni atoms within a distance of 
3.0 {\AA}.  
Table~\ref{TABLE1} lists the average coordination number, $\langle N \rangle$, 
of each atom, as well as the average number $\langle N_{\rm Ni}\rangle$, 
$\langle N_{\rm Pd}\rangle$, and $\langle N_{\rm P}\rangle$  
of Ni, Pd, and P, respectively. 
An examination of Figs.\,\ref{rdfplot}(b) and \ref{rdfplot}(c) suggests 
that Pd and P has the highest and lowest coordination numbers, 
respectively. Of all the different bonds formed by Pd, the Pd--Pd bond 
is the most dominant, followed by Pd--Ni, and Pd--P.  In 
the presence of virtually no P--P bonds, P atoms are 
segregated in the network from each other.  This 
observation has been noted by Kumar~{\it et al}.\cite{Kumar2011}

\begin{table}
\caption{\label{TABLE1}
Average coordination number each atom type: $\langle N \rangle$ is the total 
average coordination for a given atom type, $\langle N_{\rm X}\rangle$ is 
average number atom of type X (X=Ni, Pd, P) bonded to a given atom type. 
(Note that, for  a given atom type, $\langle N \rangle =  \langle N_{\rm Ni}\rangle 
+  \langle N_{\rm Pd}\rangle +  \langle N_{\rm P}\rangle$)
}
\begin{ruledtabular}
  \begin{tabular}{ccccc}
   Atom type & $\langle N \rangle$ & $\langle N_{\rm Ni}\rangle$ & $\langle N_{\rm Pd}\rangle$ & $\langle N_{\rm P}\rangle$ \\
\hline
    Ni       &       11.27         &   3.71                        &      5.20                   & 2.36     \\
    Pd       &       13.57         &   5.20                        &      6.19                   & 2.18     \\
    P        &       9.06          &   4.71                        &      4.35                   & 0.03     \\
    \end{tabular}
   \end{ruledtabular}
\end{table}

\begin{figure}[t!]
\includegraphics[clip,width=0.8\linewidth]{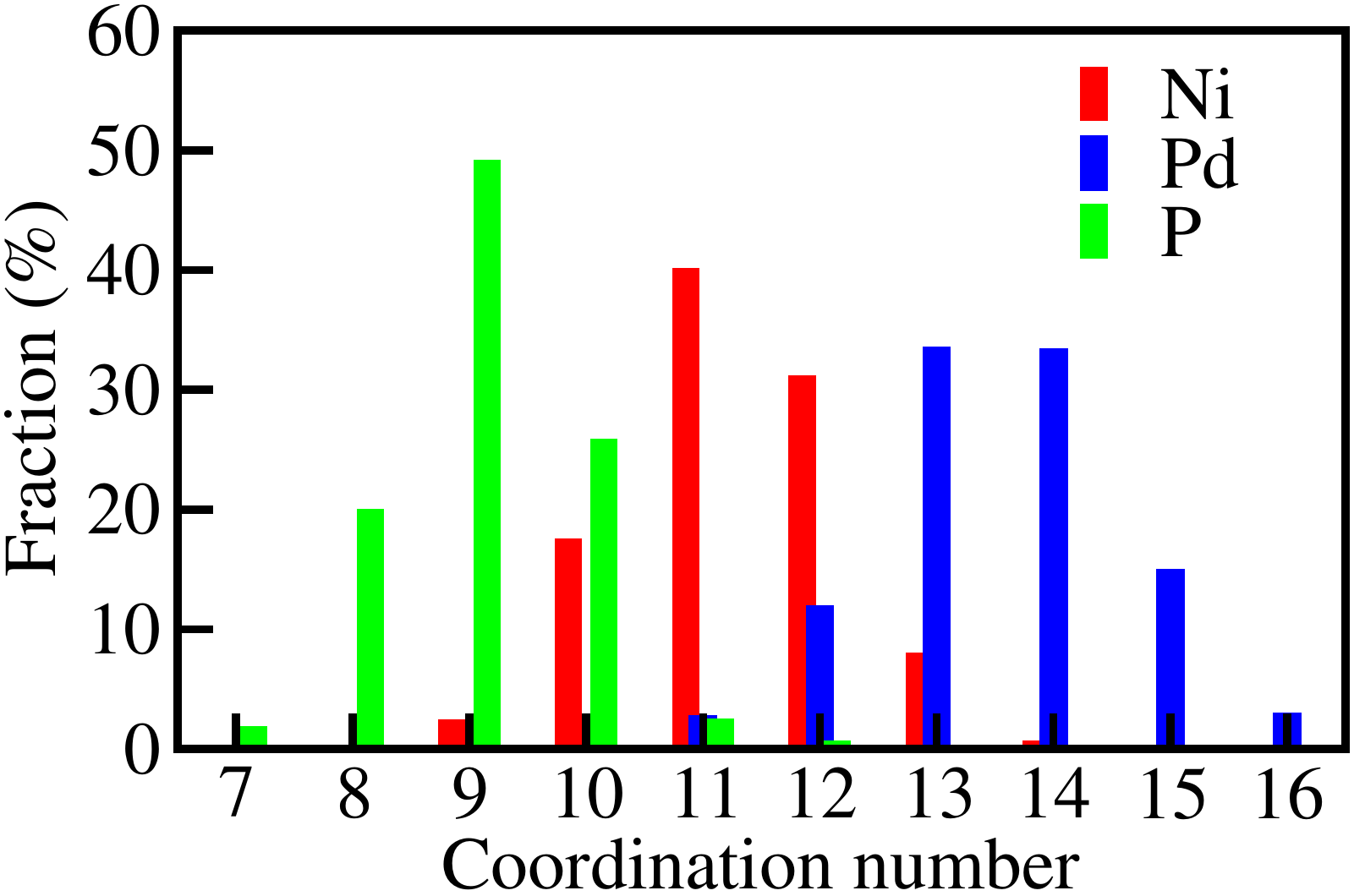}
\caption{\label{nnplot}
(Color online)
The distributions of the coordination number for each atom type, averaged 
over ten (10) independent configurations. The results for Ni, Pd, P atoms are 
shown in red, blue, and green colors, respectively. 
}
\end{figure}

Figure~\ref{nnplot} shows the distribution of the average coordination 
numbers for each atomic species.  A further breakdown of the nuances of 
the coordination numbers is presented in Table~\ref{TABLE2}.  From 
Fig.\,\ref{nnplot}, we see that the coordination numbers for Ni lies 
in the range from 9 to 15 with 40\% of the Ni atoms are 11-fold 
coordinated and 31\% are 12-fold coordinated. A breakdown of 
11-fold and 12-fold coordinated atoms for Ni in Table~\ref{TABLE2} (top) 
suggests Ni mostly forms bonds with Pd (45\%-47\%) -- an observation 
that agrees with $\langle N_{\rm Pd}\rangle$=5.2 being the largest 
coordination number for Ni in Table~\ref{TABLE2}. 
Likewise, the results for Pd in Fig.\,\ref{nnplot} indicates 
that the coordination numbers for Pd ranges from 10 to 17, 
with the dominant coordination numbers being 13 and 14. 
Pd atoms largely form bonds with another Pd atoms.  
A similar analysis confirms that nearly 50\% of P 
atoms are 9-fold coordinated, with the remainder 
being 8 and 10-fold coordinated. An examination of 
these 8-, 9-, and 10-fold coordinated P atoms in 
Table~\ref{TABLE2} shows that P atoms prefers 
to bond with Pd and Ni atoms (approximately 50\% and 
46--49\%, respectively). Table \ref{TABLE2} for P atoms 
show only 0.3\% P-P bonds have been realized in 
the network.  In summary, the bonding trend between atom types 
observed in this study is consistent with the results from previous {\it ab initio} 
studies based on small models.\cite{Kumar2011, Reyes2015}

\begin{table}
\caption{\label{TABLE2}
Statistics of coordination numbers (N) for different atomic species.  
$\langle f \rangle$ denotes fraction of $N$-fold coordinated atom 
for the atom species under consideration (vertical axis 
in Fig.\,\ref{nnplot}). For an $N$-fold coordinated atom, 
columns 3, 4, and 5 denote the distribution of atoms 
(in \%) bonded to the central atom. 
}
\begin{ruledtabular}
  \begin{tabular}{ccccc}
      \multicolumn{5}{c}{Ni}\\
      \hline
	         $N$ & $\langle f \rangle$     & bonded to Ni  & bonded to Pd &  bonded to P \\
		    &       (\%)              &     (\%)      &   (\%)       &   (\%)       \\
      \hline
	         9  & 2.4                     & 22.6               & 56.3           & 21.1\\
	         10 & 17.5                    & 28.3               & 50.4           & 21.3\\
	         11 & 40.1                    & 32.1               & 47.0           & 20.9\\
	         12 & 31.2                    & 34.4               & 45.1           & 20.5\\
	         13 & 8.0                     & 40.4               & 38.3           & 21.3\\
	         14 & 0.7                     & 42.9               & 33.9           & 23.2\\
	         15 & 0.1                     & 46.7               & 26.7           & 26.7\\
	         all & 100                    & 33.0               & 46.1          & 20.9\\
      \hline
      \multicolumn{5}{c}{}\\
      \multicolumn{5}{c}{}\\
      \hline
      \hline
      \multicolumn{5}{c}{Pd}\\
      \hline
	         N & $\langle f \rangle$     & bonded to Ni  & bonded to Pd &  bonded to P \\
		    &       (\%)              &     (\%)      &   (\%)       &   (\%)       \\
      \hline
	         10 & 0.2                     & 10.0               & 80.0           & 21.1\\
	         11 & 2.8                     & 30.0               & 57.6           & 21.3\\
	         12 & 11.9                    & 34.6               & 51.0           & 20.9\\
	         13 & 33.5                    & 35.7               & 48.7           & 20.5\\
	         14 & 33.3                    & 39.6               & 44.5           & 21.3\\
	         15 & 14.9                    & 42.0               & 40.2           & 23.2\\
	         16 & 2.9                     & 47.7               & 32.7           & 26.7\\
	         17 & 0.5                     & 54.9               & 23.5           & 26.7\\
	         all & 100                    & 38.3               & 45.6           & 16.1\\
      \hline
      \multicolumn{5}{c}{}\\
      \multicolumn{5}{c}{}\\
      \hline
      \hline
      \multicolumn{5}{c}{P}\\
      \hline
	         N & $\langle f \rangle$ & bonded to Ni  & bonded to Pd &  bonded to P \\
		    &       (\%)          &     (\%)      &   (\%)       &   (\%)       \\
      \hline
	         7 & 1.8                      & 57.1               & 42.7           & 0\\
	         8 & 20.0                     & 50.9               & 49.1           & 0\\
	         9 & 49.2                     & 50.6               & 49.3           & 0.2\\
	         10 & 25.8                    & 53.4               & 45.9           & 0.7\\
	         11 & 2.5                     & 60.0               & 38.2           & 1.8\\
	         12 & 0.7                     & 52.1               & 47.9           & 0\\
	         all & 100                    & 51.8               & 47.9           & 0.3\\
    \end{tabular}
   \end{ruledtabular}
\end{table}

\begin{figure}[t!]
\includegraphics[clip,width=\linewidth]{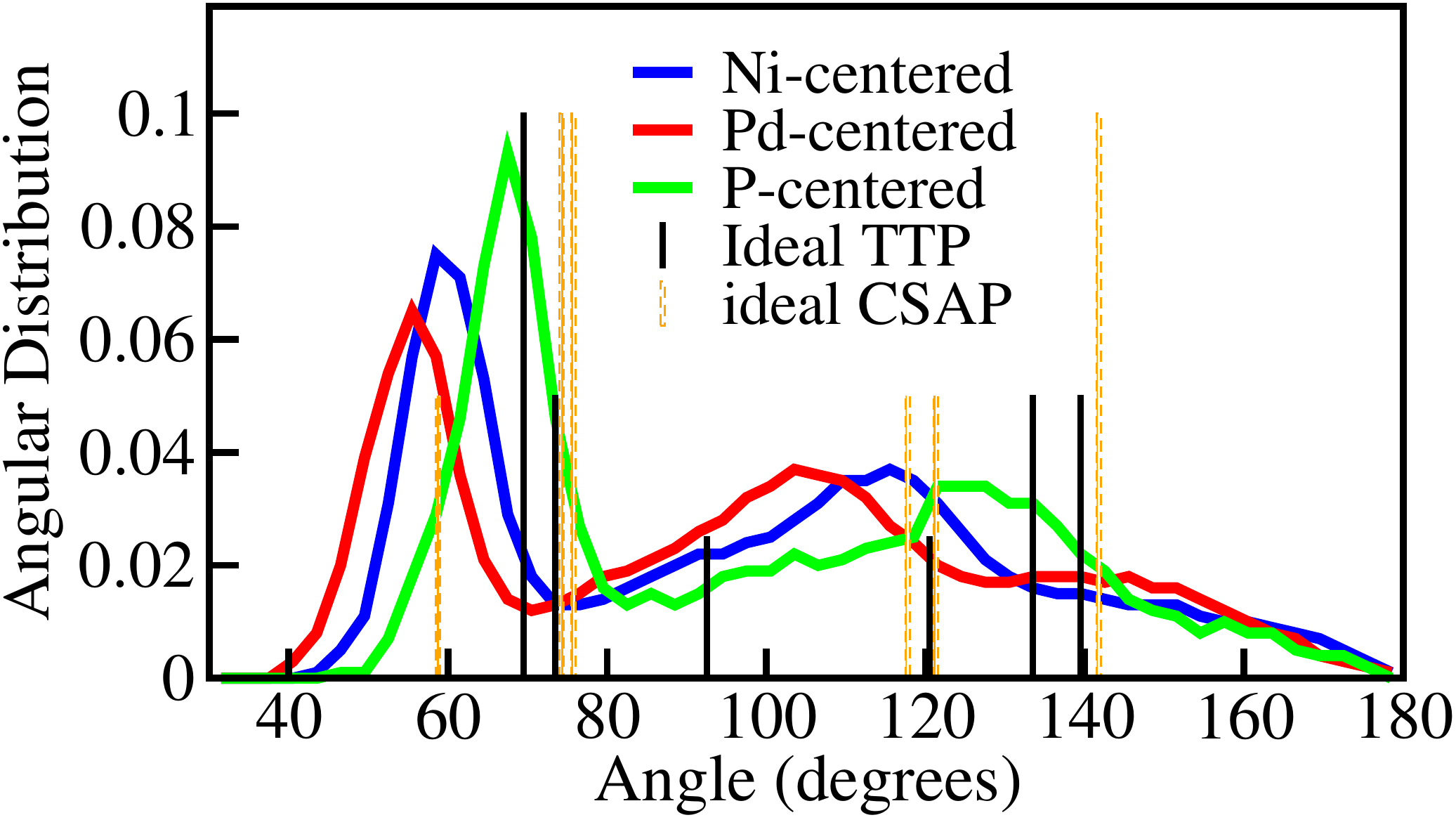}
\caption{\label{bond_angle}
(Color online)
Configuration-averaged bond-angle distributions for Ni (blue), Pd (red), 
and P (green) centers.  The vertical lines denote the bond-angle 
distributions in an ideal trigonal tricapped prism (TTP) (solid black 
vertical lines) and an ideal capped square anti-prism (CSAP) (dashed yellow 
vertical lines). 
}
\end{figure}

Figure \ref{bond_angle} presents the atom-centered bond-angle 
distributions for Ni, Pd, and P. The distributions show a bi-modal 
character with peaks in the vicinity of 60{\deg}$\pm$10{\deg} and 120{\deg}$\pm$10\deg. 
The peak positions for the P-centered bond-angle 
distribution are particularly noteworthy because they bear close 
resemblance to the angular distribution of water molecules 
around +3 lanthanide and actinide metal ions.\cite{Atta-Fynn2011} 
In +3 lanthanide and actinide ions, the geometry of water 
molecules with an angular distribution similar to the P-centered 
bond-angle distribution takes the form of tricapped trigonal prism (TTP) and 
capped square anti-prism (CSAP), both correspond to a 
metal coordination of 9. The bond-angle distributions of 
ideal TTP and CSAP geometries are shown in Fig.\,\ref{bond_angle} 
as a palisade of $\delta$-functions. 
It is evident that the peaks in the P-centered bond-angle distribution 
closely follow the center of the  $\delta$-functions from 
ideal TTP and CSAP structures, indicating the presence of distorted 
TTP and CSAP structural units in {\PDNI} glass. To examine 
this further, we performed a search for approximate TTP and CSAP 
structural units, centered on P, in the network and found multiple 
occurrences of such units. 

\begin{figure}[t!]
\centering
\subfigure[~Ideal TTP]{\includegraphics[width=0.4\linewidth]{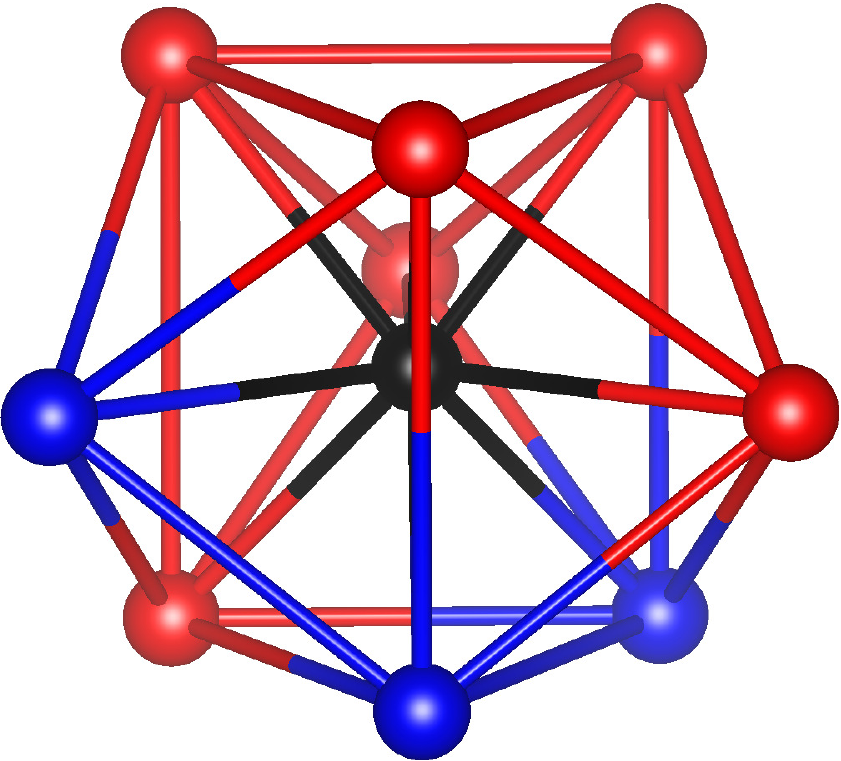}\label{ttpa}}
\subfigure[~Simulated TTP]{\includegraphics[width=0.4\linewidth]{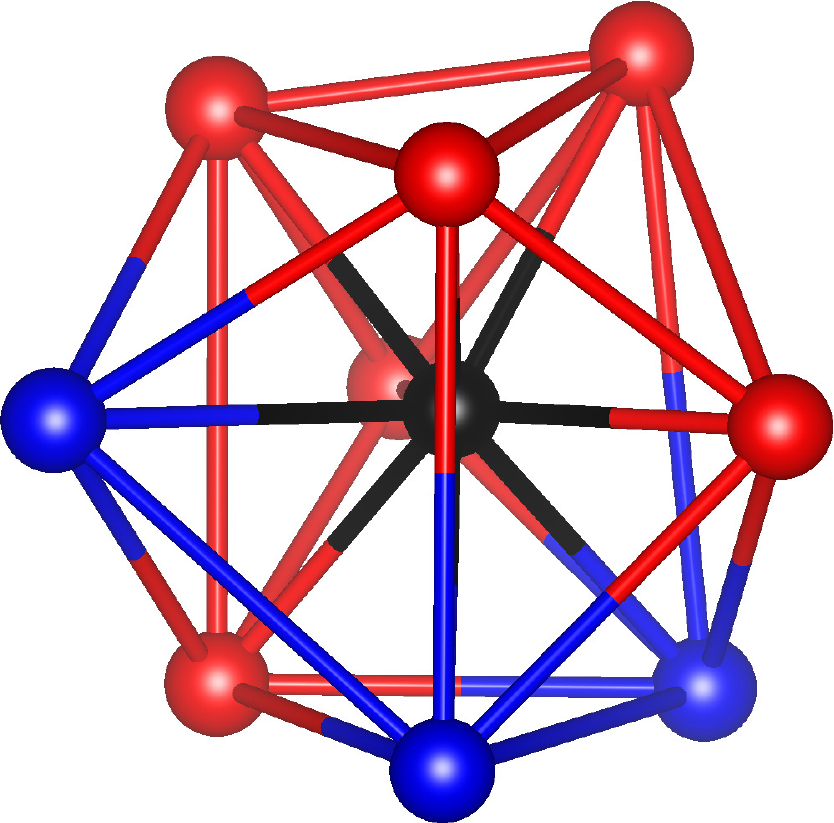}\label{ttb}}
\subfigure[~ideal CSAP]{\includegraphics[width=0.45\linewidth]{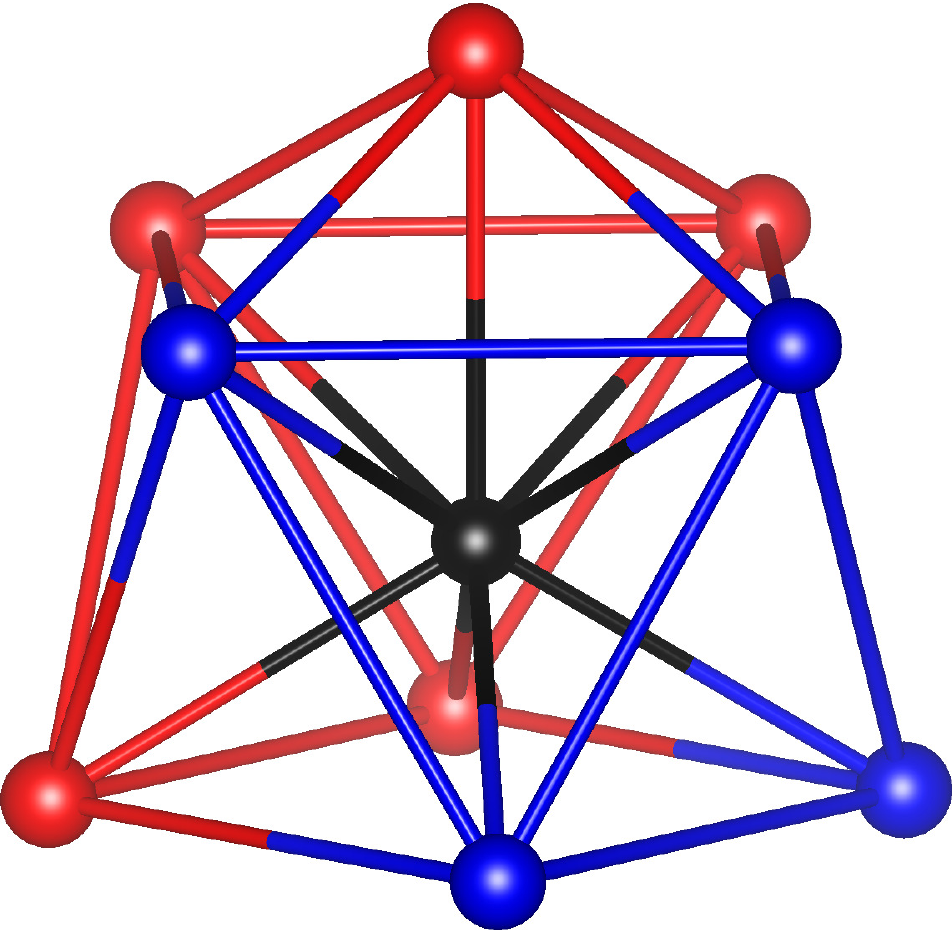}\label{csapa}}
\subfigure[~Simulated CSAP]{\includegraphics[width=0.4\linewidth]{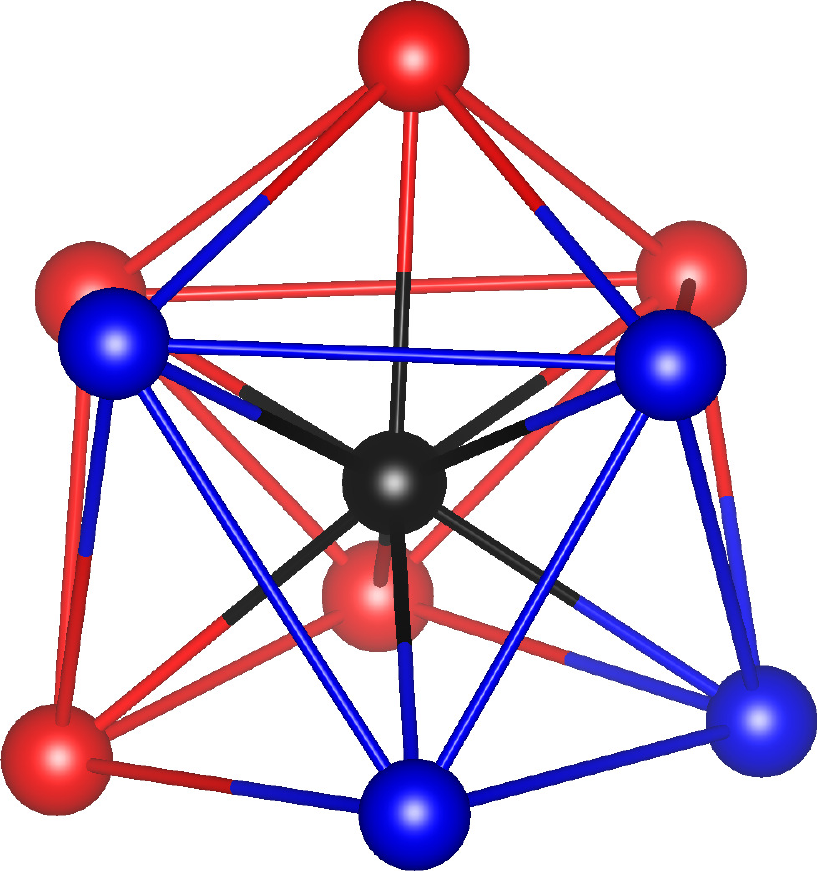}\label{csapb}}
\caption{\label{clusters}
(Color online) (a) and (b): Comparison of P-centered ideal tricapped trigonal 
prism (TTP) geometry with a distorted version from simulations. 
(c) and (d):  Comparison of P-centered ideal capped square anti-prism (CSAP) 
geometry geometry with a distorted version from simulation. Pd, Ni, and P 
atoms are shown in blue, red, and black colors, respectively.}
\end{figure}

In Fig.\,\ref{clusters}, we have shown examples of TTP and CSAP 
clusters found within {\PDNI} networks along with their ideal 
counterparts. A description of ideal TTP and CSAP clusters can be 
found in Ref.\;\onlinecite{Atta-Fynn2011}. Although the clusters 
found in {\PDNI} glassy networks exhibit noticeable 
distortions, the resemblance to the ideal TTP and CSAP 
structures is clearly visible in Fig.\,\ref{clusters}. 
These distortions can be largely attributed to the 
broadening of the angular distributions in glassy {\PDNI} 
environment but contributions from finite-temperature 
effects are also expected to play a part.
The existence of such structural units was reported by 
Kumar {\etal}~\cite{Kumar2011} in {\it ab initio} 
studies of {\PDNI} glass and by Rainer {\etal}~\cite{Rainer2006} 
and in phosphide-based metal clusters calculations. 
A characteristic feature of TTP and CSAP structures is that 
they can easily transform from one to other under the 
influence of temperature. It has been observed that an 
array of coordination structures exists that are closely 
related to the TTP and CSAP structures.  For example, 
an 8-fold coordinated atom is usually a TTP or a 
CSAP structure with a missing capping atom; a 10-fold coordinated 
atom could be a bicapped square anti-prism (CSAP structure 
plus an additional capping atom); a 12-fold coordinated 
atom is a quadruple-capped square anti-prism (CSAP structure 
plus three additional capping atoms). TTP, CSAP, and their 
low and high coordination geometric derivatives are expected 
be the building blocks of on the nanometer-scale structural 
order in this BMG. However, given the size of the 
supercell of about 8 {\AA} containing only 300 atoms, it is difficult for 
the system to form structural order on the nanometer 
length scale.  

\begin{figure}
\includegraphics[clip,width=0.75\linewidth]{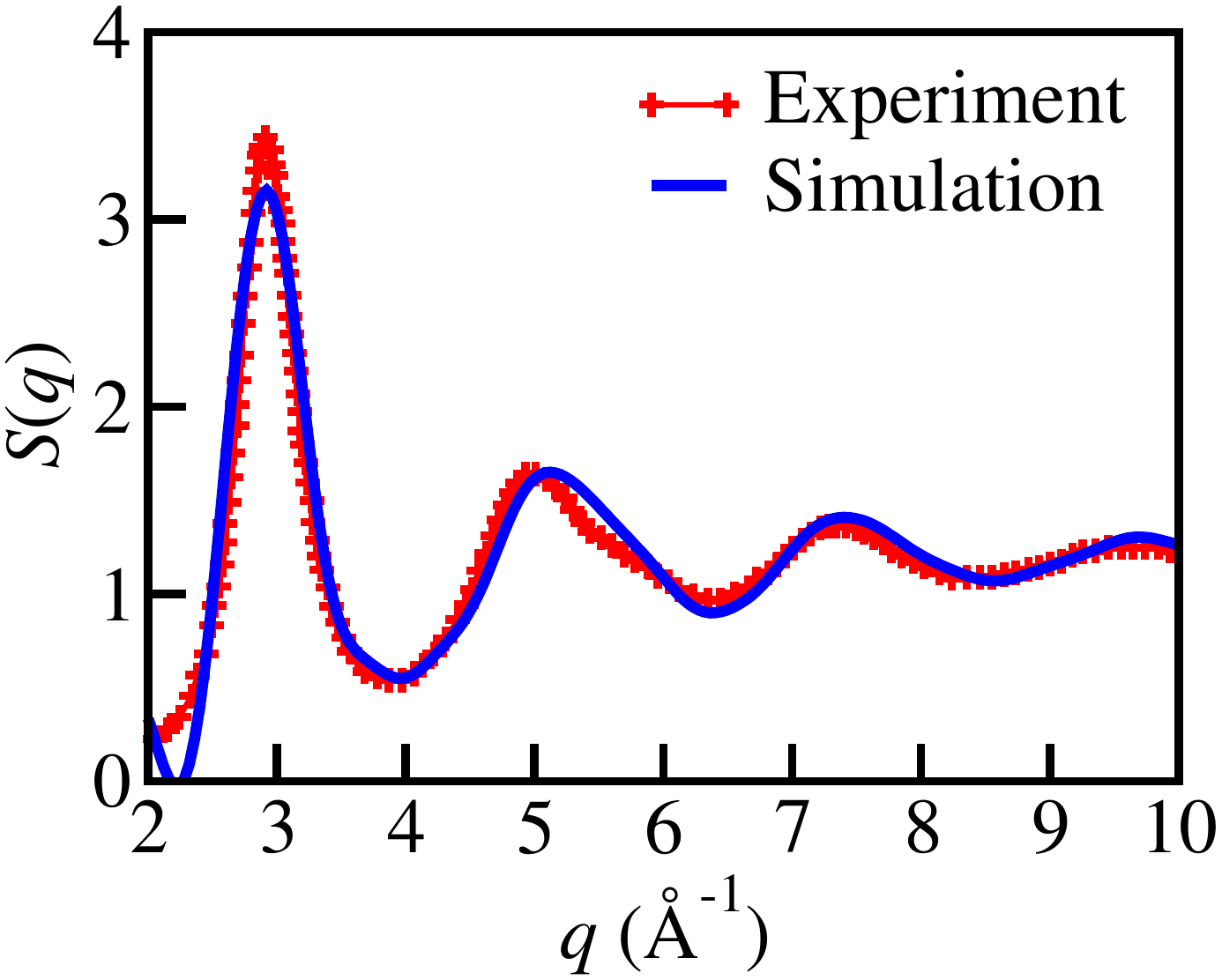}
\caption{\label{structure_factor}
(Color online)
The static structure factors of {\PDNI} from AIMD simulations at 300 K 
and experiments. Experimental data are from differential scanning 
calorimetry measurements at 313 K due to Lan {\etal}.\cite{La2017}
}
\end{figure}

Figure~\ref{structure_factor} depicts the simulated static 
structure factor, averaged over 10 configurations, along 
with the experimental structure-factor data from a recent 
differential scanning calorimetry (DSC) experiments, due 
to Lan~{\etal}\cite{La2017} The static structure factor, 
$S(q)$,  can be obtained directly from the position of 
the atoms in the network: 
\be
S(q) = \frac{\left\langle\sum\limits_{j=1}^N \sum\limits_{k=1}^N
f_{j}(q)f_{k}(q)\exp\left(-\imath {\bf q} \cdot {\bf r}_{jk}\right) \right\rangle}
{\sum\limits_{j=1}^N {\left(f_{j}(q)\right)}^2},
\label{sk-eq}
\ee
\noindent where ${\bf r}_{jk}={\bf r}_{k} - {\bf r}_{j}$ is the vector from the 
position ${\bf r}_{j}$ of the $j$-th atom to the position ${\bf r}_{k}$ of the $k$-th atom 
(with periodic boundary conditions taken into account to mimic bulk effects), $f_j$ is the atomic 
form factor of the $j$-th atom, ${\bf q}$ is wave-vector transfer of magnitude  $q$ and symbol 
$\langle . \rangle$ indicates the rotational averaging of ${\bf q}$ of magnitude $q$ over a solid angle
of $4\pi$\cite{averaging} and averaging over all 10  configurations. 
The atomic form factor was computed using the Gaussian representation: 
\be
f_{j}(q)=\sum\limits_{j=1}^4 a_j\exp\left(-b_j\left(\frac{q}{4\pi}\right)^2\right) + c, 
\ee
\noindent where $a_j$, $b_j$, and $c$ are atom-type specific empirical constants taken from 
Ref.\;\onlinecite{Brown2004}. The match between the simulated structure factor with
experiment in Fig.\,\ref{structure_factor} is very good, further validating the accuracy 
of the structural models. The simulated data is also in good agreement with the experimental 
data by Egami {\it et al}.\cite{Egami1998}

\subsection{Comparison of Simulated and Experimental XAFS Spectra}
To further verify the credibility of the models presented in this work, we compare a key structural property, 
namely the extended X-ray absorption fine-structure spectroscopy (EXAFS), 
of our simulated atomistic configurations to available experimental data. For a given configuration, 6 {\AA} clusters 
centered on the absorber (that is the Pd or K atom) 
are carved out used as input to the FEFF9 {\it ab initio} multiple scattering code,\cite{Rehr2000,Rehr2009} 
which produces an $K$ edge EXAFS spectrum for each cluster.
The separate spectra for the individual clusters are then averaged to yield 
EXAFS spectrum for that configuration. The EXAFS spectra for the ten (10) atomistic 
configurations was then averaged to obtain to representative EXAFS spectrum 
$\chi(k)$, of Pd and Ni $K$-edges. Depicted in Figures are the $k^2\chi(k)$ EXAFS 
spectra and its corresponding Fourier transform ${\bar\chi}(R)$ given by:

\begin{equation}
   {\bar\chi}(R)=\frac{1}{2\pi}\int_{0}^{\infty}k^2\chi(k)\Omega(k)e^{i2kR}
   \label{fft_chi}
\end{equation}

\noindent where $\Omega(k)$ is the Hanning window function.

\begin{figure}
\centering
\subfigure[~Pd $K$-edge EXAFS]{\includegraphics[clip,width=0.49\linewidth]{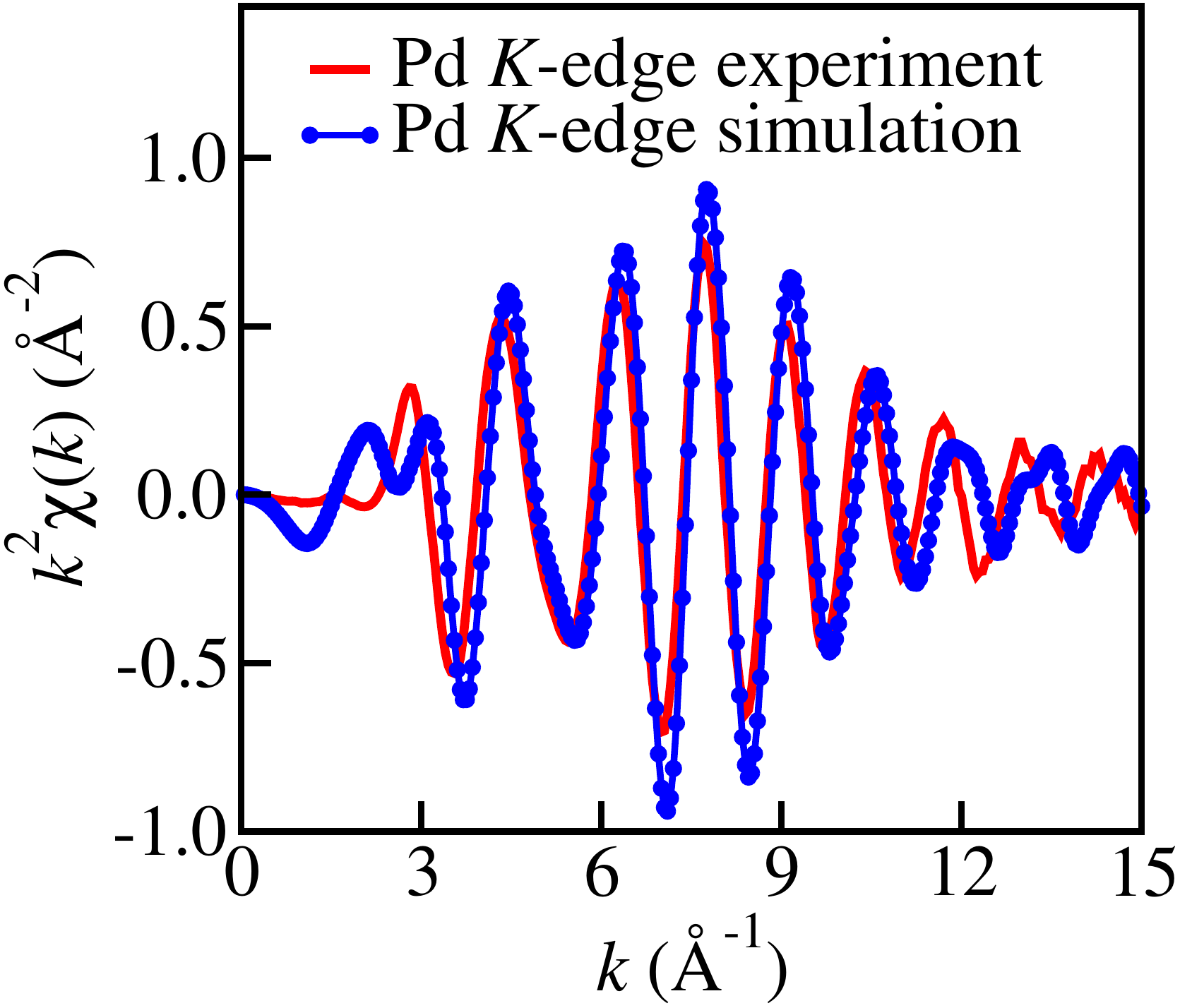}\label{Pd_xafs}}
\subfigure[~Fourier transform of plot in (a)]{\includegraphics[clip,width=0.49\linewidth]{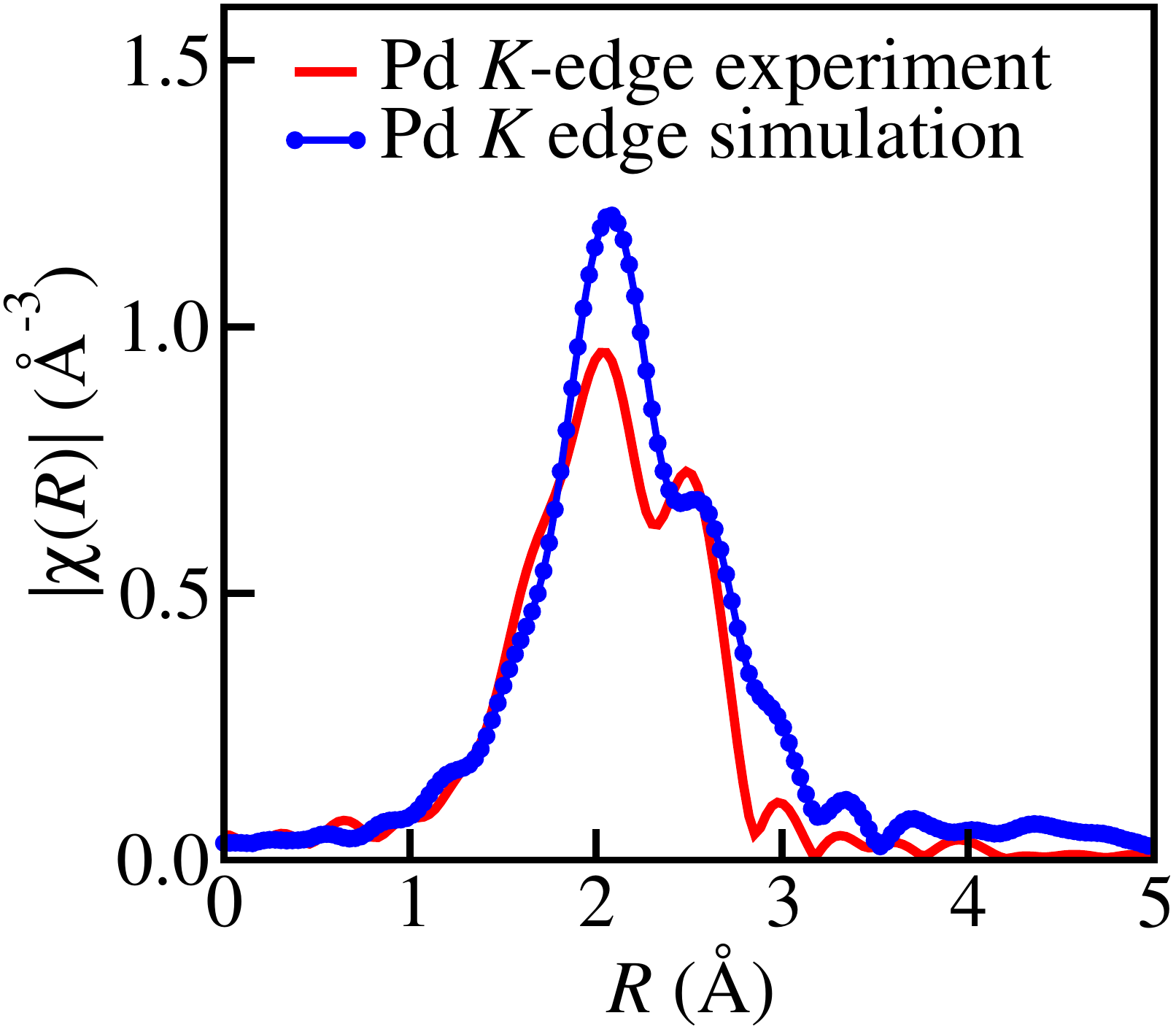}\label{Pd_FFT}}
\subfigure[~Ni $K$-edge EXAFS]{\includegraphics[clip,width=0.49\linewidth]{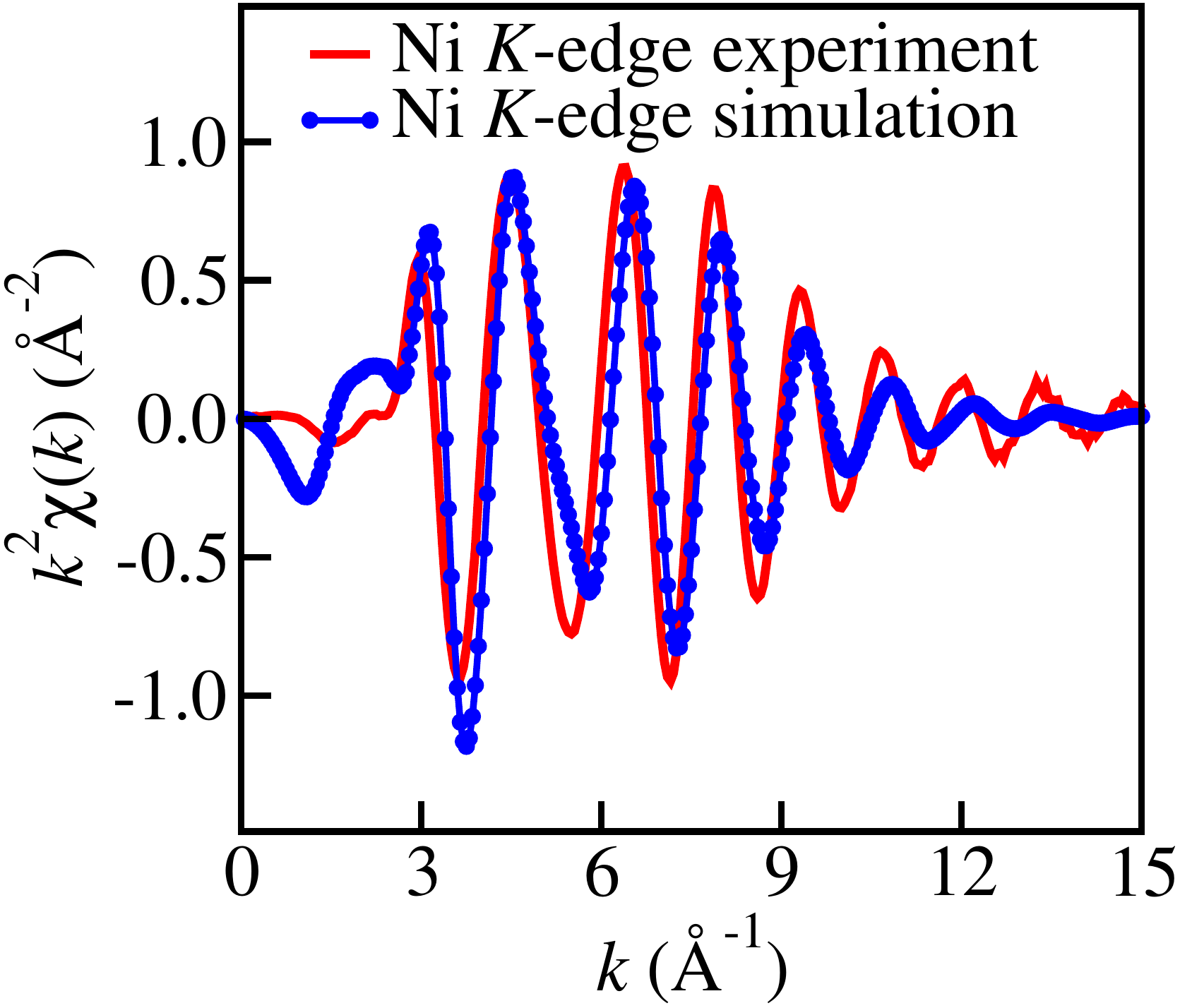}\label{Ni_xafs}}
\subfigure[~Fourier transform of plot in (c)]{\includegraphics[clip,width=0.49\linewidth]{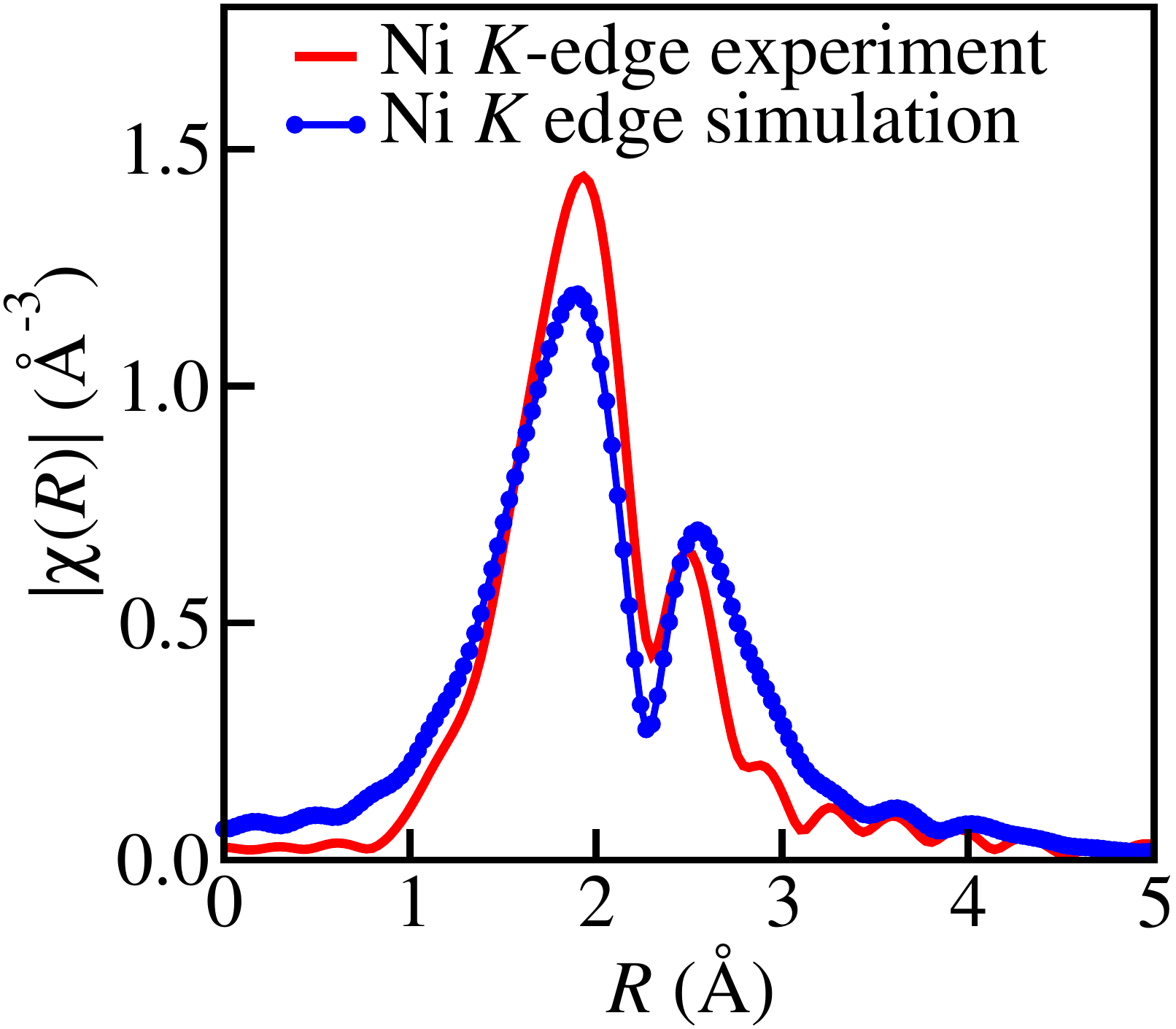}\label{Ni_FFT}}
\caption{\label{exafs}
(Color online) Comparison of the simulated Pd and Ni $K$-edge EXAFS spectra with experimental data.
The experimental data is due to Kumar and co-workers.\cite{Kumar2011}
Top: (a) $k^2\chi(k)$ EXAFS spectra of the Pd $K$-edge; (b) Magnitude, $\vert\bar{\chi}(R)\vert$, of the Fourier 
transform, of plot in (a) defined by equation~\ref{fft_chi}.
Bottom: (a) $k^2\chi(k)$ EXAFS spectra of the Ni $K$-edge; (b) Magnitude, $\vert\bar{\chi}(R)\vert$, of the Fourier
transform, of plot in (a) defined by equation~\ref{fft_chi}.
}
\end{figure}

In Figs.~\ref{Pd_xafs} and~\ref{Pd_FFT}, the simulated $K$-edge $k^2\chi(k)$ EXAFS 
and the magnitude of its Fourier transform, $\vert\bar{\chi}(R)\vert$ for Pd  
are depicted and compared recent experimental data due to Kumar {\it et al}.
\cite{Kumar2011}  In Fig.\,\ref{Pd_xafs}, we observe that the
frequency of oscillations and amplitude of the simulated spectra matches fairly well with the experimental 
spectra. The Fourier-transformed data is displayed in the bottom
Figure. The Figure does not include the correction for the phase-shift 
associated with the absorber--scatterer interactions and hence the displayed radial distances are
lower than the true radial distances in the partial RDF for Pd. The plots for Ni $K$-edge 
in Fig.\,\ref{Ni_xafs} shows amplitudes in the simulated data which match fairly well with 
experimental data. The Fourier-transformed data for the Ni $K$-edge in Fig.\,\ref{Ni_FFT} also 
correlates well with the experimental data. 

%
%

\subsection{Electronic properties}
\begin{figure}[ht]
\includegraphics[clip,width=0.75\linewidth]{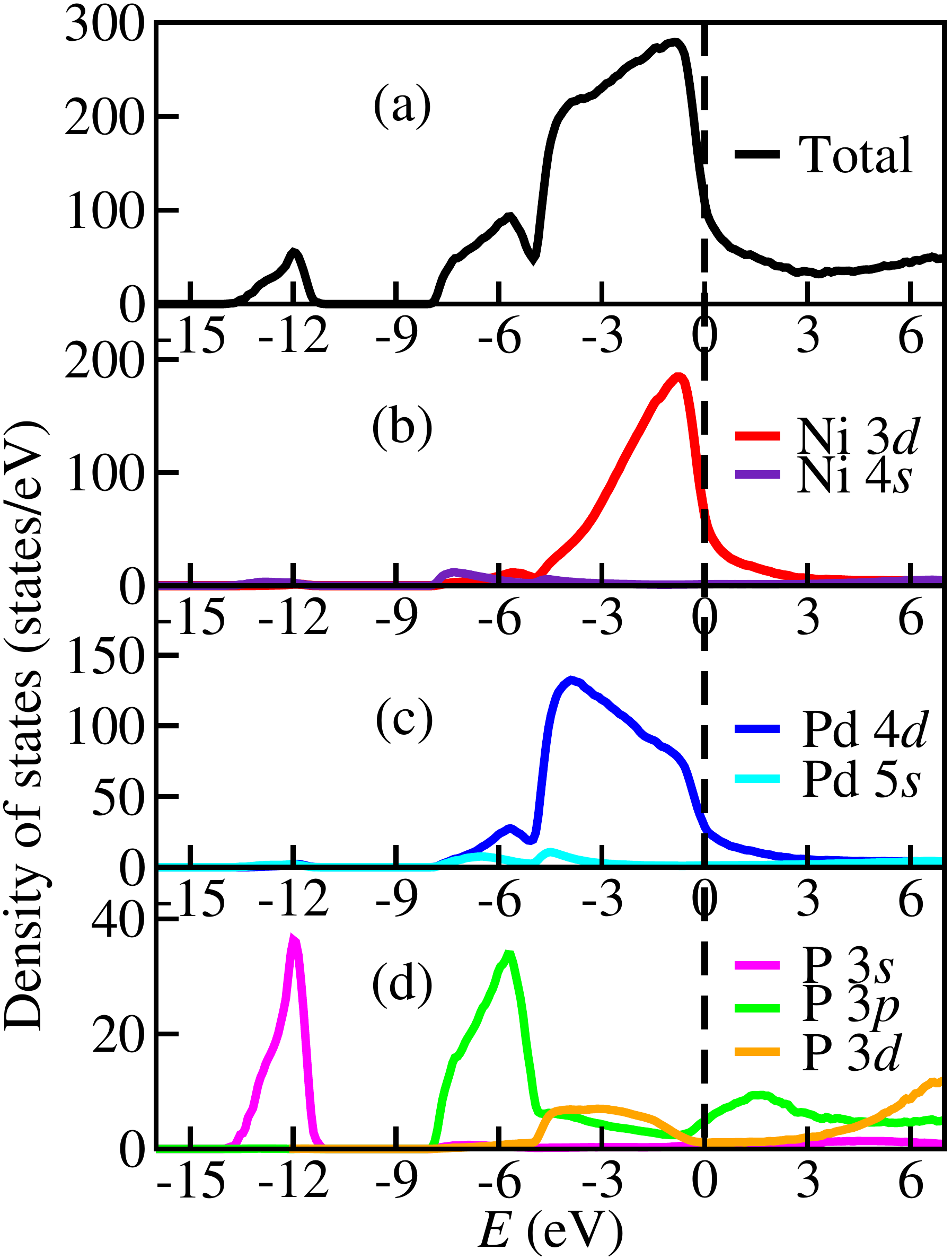}
\caption
{
\label{density_of_states}
(Color online)
The electronic densities of states (EDOS) from a 300-atom model 
of {\PDNI}: (a) total EDOS; (b) EDOS for 
Ni atoms projected onto the 3$d$ and 4$s$ orbitals; (c) DOS for 
Pd atoms projected onto the 4$d$ and 5$s$ orbitals; (d) DOS for 
P atoms projected on the $3p$ and $3d$ orbitals. The Fermi 
level, indicated as a  dashed vertical line, is located 
at $E$=0 eV. 
}
\end{figure}

Depicted in Fig.\,\ref{density_of_states} are the total and angular-momentum resolved (for each atom 
type) electronic density of states (DOS) of the single particle Kohn-Sham energy eigenvalues. 
The Fermi level was set to $E$=0 (dashed vertical line). The region between $E=-15$ eV and $E=-11.5$ eV 
is dominated by the P 3$s$ states. The P 3$s$ hybridizes with the Ni 4$s$ and Pd 5$s$ electron states 
corresponding to the Ni and Pd atoms located at the vertices of the TTP and CSAP geometries.\cite{Takeuchi2007,Guan2012}
The region between $E=-9$ eV and $E=-5$ eV is dominated by the P 3$p$ states, with a 
small admixture by the Ni and Pd $s$ and $d$ states; these 
bands are responsible for the capping bonds between P and the capping Ni and Pd atoms in the 
TTP and CSAP structures. The strong hybridization between the P electrons states and the 
Ni and Pd states suggest the formation of P--Ni and P--Pd covalent bonds. The region between 
$E=-5$ eV and the Fermi level is dominated by the Ni 3$d$ and Pd 4$d$ bands, with a small 
contribution from the P 3$d$ band. This energy region is responsible for the Pd--Pd, Pd--Ni, 
and Ni--Ni metallic bonding. Overall, the DOS analysis indicate that the P--Ni and P--Pd 
bonds are much stronger than metallic bonds, and hence the presence of P makes the 
BMG more cohesive.

\subsection{Vibrational properties}
The vibrational analysis of our models was based on the vibrational  eigenfrequencies  and
eigenvectors obtained by diagonalizing the dynamical matrix. Specifically, we computed the 
atom-projected vibrational density of states (VDOS), total VDOS, generalized VDOS, and the 
inverse participation ratio (IPR).
 
The vibrational density of states, $g_\alpha(\omega)$, projected on atoms of type $\alpha$ at frequency $\omega$ 
is given by 
\be
g_\alpha(\omega)=\sum\limits_{j=1}^{N_\alpha}\sum\limits_{i=1}^{3N}\lvert \chi_{ij}\rvert^{2}\delta(\omega-\omega_i)
\ee 
\noindent where $\alpha$=Ni, Pd or P, $\{\omega_i:\,i=1,3N\}$ are the phonon eigenfrequencies, $N_\alpha$ is the total number of atoms of type 
$\alpha$, and $\lvert \chi_{ij}\rvert^{2}$ is the contribution of atom $j$ to the square of the amplitude of the 
normalized phonon eigenvector corresponding to the $i^{\rm th}$ phonon eigenfrequency $\omega_i$. 
From the definition of $g_\alpha(\omega)$, the total vibrational density of states, $g(\omega)$, inverse 
participation ratio, ${\rm IPR}(\omega)$, and {\it generalized vibrational density of states} (GVDOS)\cite{Suck2002}, 
$G(\omega)$, at frequency $\omega$ are given by: 

\begin{align}
g(\omega)&=\sum\limits_{\alpha}g_\alpha(\omega)=\sum\limits_{i=1}^{3N}\delta(\omega-\omega_i)\label{eqtvdos}\\
{\rm IPR}(\omega_i)&=\sum\limits_{j=1}^{N}\lvert \chi_{ij}\rvert^{4}\label{eqipr}\\
G(\omega)&=\frac{\sum\limits_{\alpha}w_\alpha g_\alpha(\omega)}{\sum\limits_{\alpha}w_\alpha}\label{eqgvdos}\\
w_\alpha&=\frac{\exp(-2W_\alpha)c_\alpha\sigma_\alpha}{M_\alpha}\label{eqdw}
\end{align}

\noindent where $N$ is the total number of atoms, $\exp(-2W_\alpha)$ is the Debye-Waller factor for the 
atom of type $\alpha$, $c_\alpha$ is the concentration of atoms of type $\alpha$ in the system, $\sigma_\alpha$ 
is the total scattering cross-section of atom $\alpha$ and $M_\alpha$ is the atomic mass of atom of type $\alpha$. 
For this work, the weight factors $w_\alpha$ (in equations~\ref{eqgvdos} and~\ref{eqdw}) 
from the experimental work of Suck\cite{Suck2002}, namely $w_{\rm Ni}$ = 0.768, $w_{\rm Pd}$ = 0.102, 
$w_{\rm P}$ = 0.13, were employed.

\begin{figure}[ht]
\includegraphics[clip,width=0.75\linewidth]{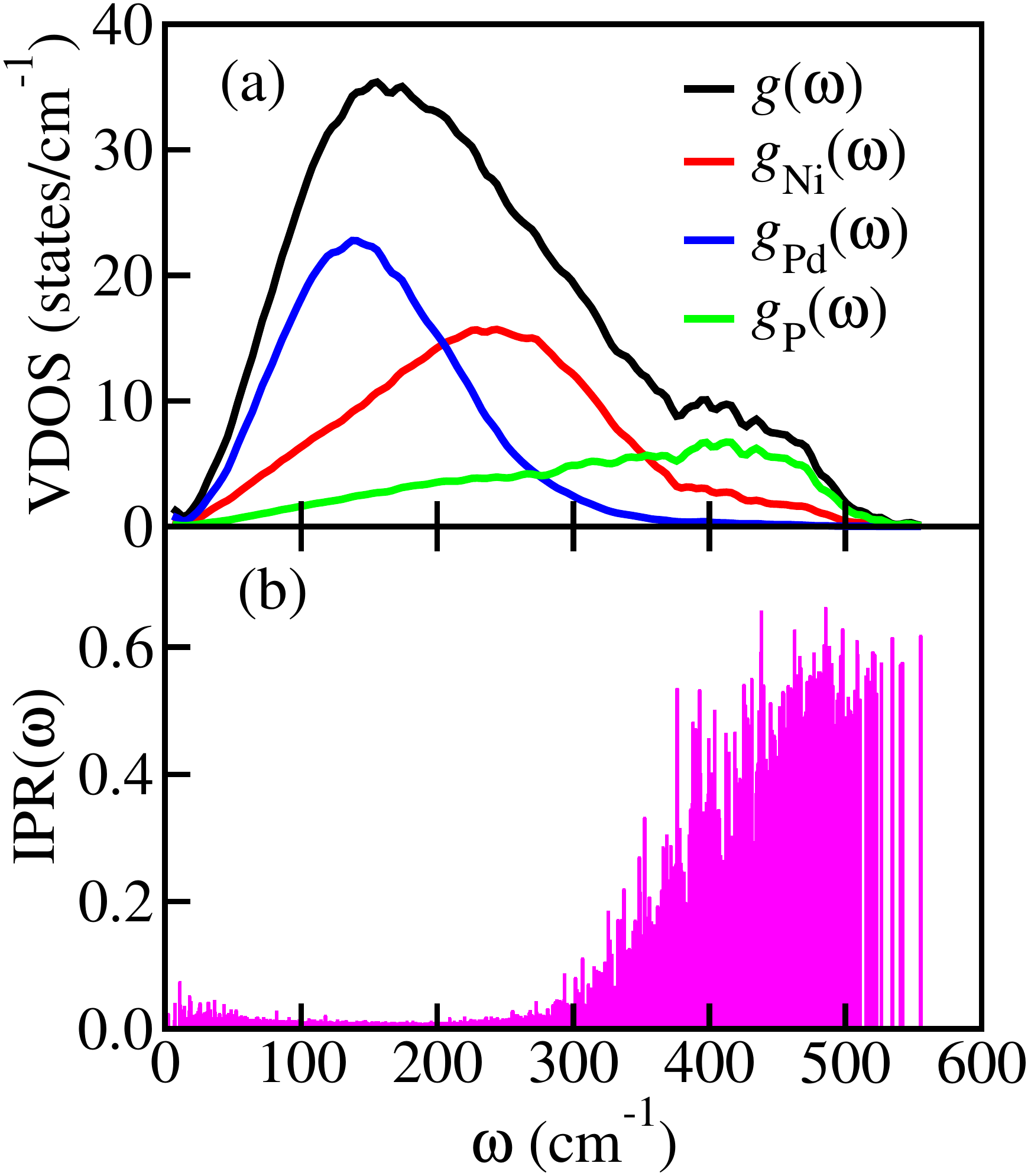}
\caption{\label{vdos_fig}
(Color online) 
Figure depicting: (a) the computed total and atom projected vibrational density of states (VDOS); 
(b) IPR as a function the eigenfrequencies.}
\end{figure}

\begin{figure}[ht]
\includegraphics[width=0.6\linewidth]{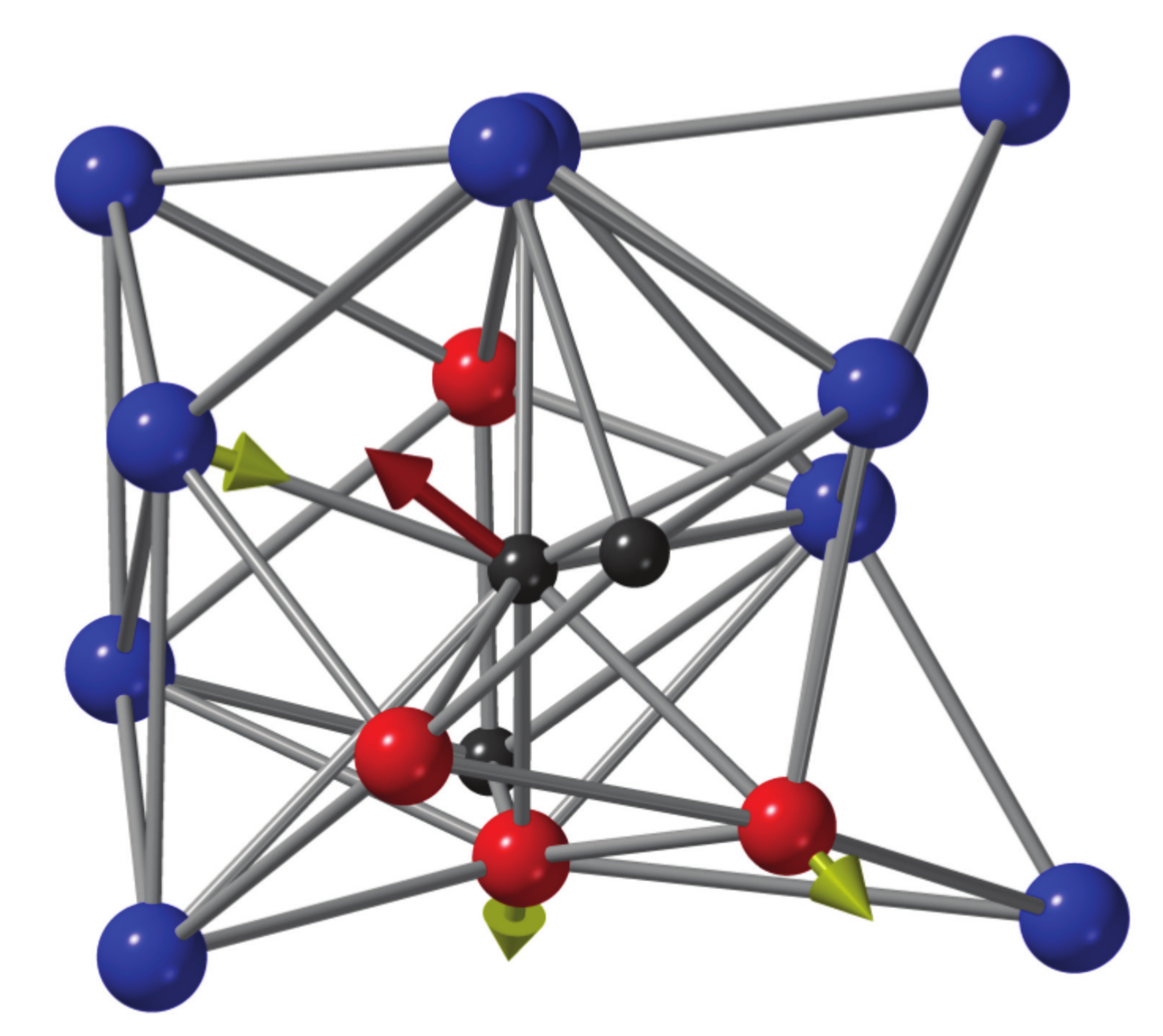}
\caption{\label{mode}
(Color online) 
An example of a localized mode in {\PDNI} with IPR = 0.66 and $\omega=485\, {\rm cm}^{-1}$. 
The red arrow represents an 80\% contribution by P (black) to 
the vibrational amplitude while the yellow arrows 
represent approximately 6\% contribution each from two Ni (red) atoms 
and one Pd (blue) atom. The average bond length between the central 
P atom and the neighboring three atoms is about 2.2 {\AA}, which is 
indicative of the vibrational localization length of the mode. 
}
\end{figure}

\begin{figure}[ht]
\includegraphics[clip,width=0.75\linewidth]{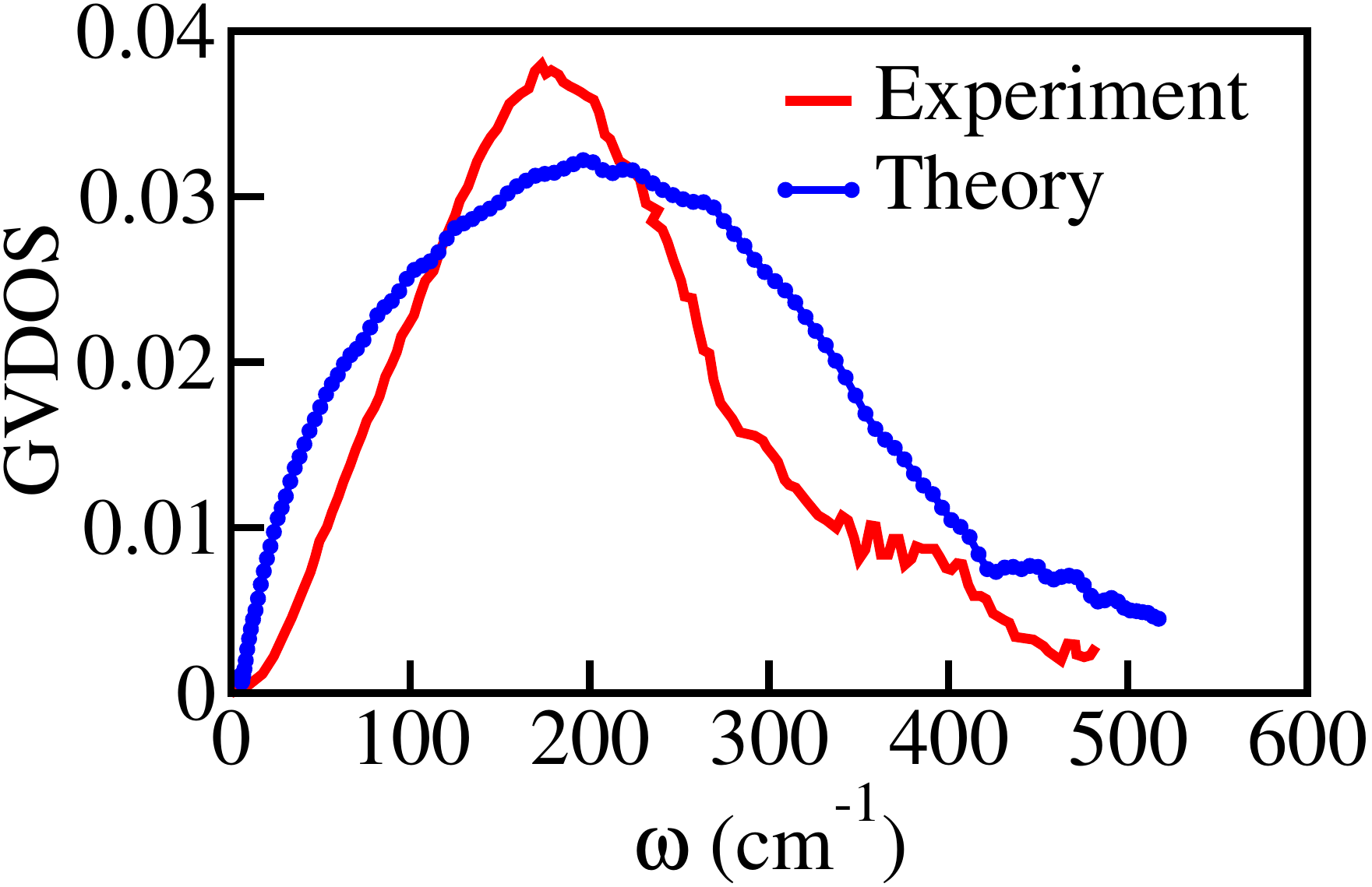}
\caption{\label{gvdos_fig}
(Color online) Comparison of the calculated generalized vibrational densities of states (GVDOS) with 
inelastic neutron-scattering experimental data, due to Suck.\cite{Suck2002}
}
\end{figure}

In Fig.\,\ref{vdos_fig}(a) the atom-resolved and total vibrational density of states are shown. Clearly 
the  low  frequency  states  are  dominated by Ni and Pd while the high-frequency states originate 
from the atomic vibrations  of Ni and P. The IPR shown in Fig.\,\ref{vdos_fig}(b) is measure of the amount 
eigenmode amplitude localized at the atomic sites. For a perfectly localized eigenmode conjugate to 
the eigenfrequency $\omega_i$, only a single atom gives rise to the vibrational amplitude and hence IPR($\omega_i$)=1.   
For  a perfectly delocalized eignemode corresponding to the eigenfrequency $\omega_i$, each 
atomic site makes an equal contribution to the amplitude of the eigenmode, and so IPR($\omega_i$)=1/$N$. 
From the IPR plot in Fig.\,\ref{vdos_fig}(b), an extended state-to-localized 
phonon eigenstate transition in the vicinity of $\omega=300\, {\rm cm}^{-1}$ can be observed. 
An analysis of the atomic contribution to the amplitudes of vibration of the phonon eigenmodes 
indicates that the amplitudes of vibrations of majority of the modes below $\omega=300\, {\rm cm}^{-1}$ 
are delocalized over all the atoms, while the amplitudes of the high-frequency eigenvectors were localized on P and 
about four or five neighboring Pd and/or Ni atoms.  As an example, Fig.\ref{mode} depicts the magnitude and direction of an atom's contribution to localized eigenmode amplitude 
(IPR=0.66, $\omega=485\, {\rm cm}^{-1}$). The central P atom, shown in  black color, 
makes an 80\% contribution to the amplitude, while two Ni atoms (red) and one 
Pd atom (blue) make contributions of approximately 6\% (resulting in a 
net contribution of approximately 98\% by the four atoms).
More importantly, the Ni and Pd atoms are within a radius of less than 2.3 {\AA} 
from the P atom. The nature of the mode in Fig.\,\ref{mode} appears to be 
an admixture of stretching and bending modes, centered on P, with 
the stretching modes being dominant. The other localized modes were found to 
behave in a fashion similar to the one in Fig.\,\ref{mode}. 
The small localization radius (centered on P) is a characteristic feature of 
all the localized eigenmodes.  In Fig.\,\ref{gvdos_fig}, the GVDOS is computed and compared with experimental data. 
Ni vibrations dominate the lower energy of the GVDOS, while P vibrations 
dominate the higher energy bands. Apart from the small deviations 
of the simulated GVDOS from the experimentally 
measured data, the overall match is good, further validating the reliability of the applied theoretical model.  

\section{Conclusions}

The structural, electronic, and vibrational properties of {\PDNI} bulk
metallic glass have been studied using {\it ab initio} 
molecular-dynamics simulations and total-energy minimizations.
Since multi-component BMGs are characterized by a variety of atomic
environments, an ensemble of ten (10) independent 300-atom structural
configurations were simulated and studied in an effort to take into account the
effect of configuration fluctuations on the geometric, electronic,
and vibrational properties of glassy {\PDNI}. The average static
structure factor obtained from our simulations is found to be in
excellent agreement with experimental data.
Likewise, the simulated Ni and Pd $K$-edge EXAFS spectra, averaged over 10
independent configurations, are also observed to be in good agreement
with experimental measurements.  The structural analysis of the resulting
10 configurations shows the absence of P--P bonds, with the
segregated P atoms acting as the centers for distorted tricapped
trigonal prisms and capped square anti-prism geometries.
An analysis of the electronic density of states indicates that P atoms
form strong
covalent bonds with Ni and P at low energies, while the bonding states near
the Fermi level are dominated by metallic Ni--Ni, Ni--Pd, and Pd--Pd bonds.
A further analysis of the vibrational spectrum indicates that the low-frequency
vibrational modes are dominated by the vibrations of Ni and Pd atoms, while the high-frequency
modes primarily arise from the atomic
vibrations of P atoms. The amplitudes of the high-frequency eigenmodes
are observed to be well-localized on P atoms, which comprise an admixture
of P--Ni and P--Pd stretch modes and bending modes. The generalized vibrational
density of states is also found to agree fairly closely with the
experimental data from inelastic neutron-scattering measurements.

We conclude this section with the following observation.
A question of considerable importance that the current study could
not address is the role of medium-range order on the nanometer
length scale, which has been postulated as one of the qualities
that give rise to the unique properties of BMGs.
Although the presence of a few structural motifs in the 
glassy {\PDNI} environment is indicative of the 
medium-range order, the limited size of the models does not 
permit us to address this issue directly from our simulations. 
Nonetheless, the approach presented here can be employed 
profitably to simulate larger models on the nanometer 
length scale by using small high-quality structural 
configurations as the basic building blocks, which has the correct
short-range order and the local chemistry built into it.
In future, we shall address this building-block approach by combining 
the small structural units developed here to form a super structure/unit
and annealing the resultant the super structure through
a series of temperatures. Such an approach would be computationally
efficient and the resulting models should provide an accurate 
structural basis to address the nanoscale properties of 
bulk metallic glasses.

\begin{acknowledgments}
The work was supported by U.S. National Science Foundation
under grant numbers DMR 1507166, DMR 1507118 and DMR 1507670.
We thank Profs.\,Kumar, Fujita, and Kawazoe for making the
experimental EXAFS data available to us.  Atta-Fynn
acknowledges the Texas Advanced Computing Center (TACC) at
The University of Texas at Austin for providing HPC resources
that have contributed to the results reported in this work.
\end{acknowledgments}

\bibliography{paper.bbl}

\end {document}